\documentclass{iaus}
\usepackage{graphicx}

\title[The age of the Galaxy's thick disk] 
{The age of the Galaxy's thick disk}

\author[S. Feltzing \& T. Bensby]   
{Sofia Feltzing$^1$\thanks{SF is a Royal Swedish Academy of Sciences Research Fellow
supported by a grant from the Knut and Alice Wallenberg Foundation.
} \and Thomas Bensby$^2$
}

\affiliation{$^1$Lund Observatory, \\ Box 43,
SE-22100, Lund, sweden \\ email: {\tt sofia@astro.lu.se} \\[\affilskip]
$^2$European Southern Observatory,\\ Alonso de Cordova 3107,
  Vitacura, Casilla 19001, Santiago, Chile \\email: {\tt tbensby@eso.org} 
}

\pubyear{2008}
\volume{258}  
\pagerange{1-8}
\jname{The Ages of Stars}
\editors{A.C. Editor, B.D. Editor \& C.E. Editor, eds.}
\begin{document}

\maketitle

\begin{abstract}
  We discuss the age of the stellar disks in the solar neighbourhood.
  After reviewing the various methods for age dating we discuss
  current estimates of the age of both the thin and the thick disk.
  We present preliminary results for kinematically-selected stars that
  belong to the thin as well as the thick disk. All of these dwarf and
  sub-giant stars have been studied spectroscopically and we have
  derived both elemental abundances as well as ages for them.  A
  general conclusion is that in the solar neighbourhood, on average,
  the thick disk is older than the thin disk. However, we caution
  that the exclusion of stars with effective temperatures around
  6500\,K might result in a biased view on the full age distribution
  for the stars in the thick disk.  
\end{abstract}

\firstsection 
\section{Introduction}

\bigskip

The age of a stellar population can be determined in several ways.
For groups of stars isochrones may be fitted to the stellar sequence
in the Hertzsprung-Russell diagram (HR-diagram, compare
e.g. \cite[Schuster et al. (2006)]{sn2006}) or the luminosity function
for the white dwarfs can be fitted with cooling tracks (see e.g.
\cite[Leggett et al. (1998)]{legget98}). Ages for individual stars can
be determined from the HR-diagram (if the star is a turn-off or
sub-giant star) or by utilising relations that relate the rotation or
atmospheric activity of a star to its age (examples are given by
\cite[Barnes (2007)]{barnes2007} and \cite[Mamajek \& Hillenbrand
(2008)]{mamajek2008}). Asteroseismology provides the possibility to
constrain the stellar ages very finely. A recent example of the age
determination for a young star is given in \cite[Vauclair et
al. (2008)]{vauclair}. Finally, the age of a star can be estimated by
studying the amount of various elements present in the photosphere of
the star. In particular the amount of elements such as U and Th that
decay radioactively can be used to estimate the age. Examples of this
are given by \cite[del Peloso et al. (2005)]{dp2005}. Estimating the
age from the decay of radioactive isotopes is sometimes called {\em
  nucleocosmochronology}.

All but one of these methods, nucleocosmochronology, relies on our
understanding of stellar evolution. Some of the methods work well for
young stars. This is especially true for rotation and stellar
activity (see \cite[Mamajek (2009)]{mamajek2009} and \cite[Barnes
(2009)]{barnes2009}) whilst the determination of stellar ages using
isochrones is limited in various ways depending on the type of 
star under study. 

The isochrones give the best results for turn-off and sub-giant stars
with very poor power to differentiate between different ages on the
red giant branch. In fact, the stars on the sub-giant branch are the
most desirable tracers of the age of a particular stellar population,
\cite[Sandage et al. (2003)]{sandage}.  In particular it does not
matter if the stellar temperature is well determined or not,
\cite[Bernkopf \& Fuhrmann (2006)]{bf2006}.

However, we would argue that the power of isochrone ages mainly lies
in the {\em relative} ages -- i.e. being able to say ``star A is older
than star B and it is about this big an age difference between star A
and star B''.  Such statements and determinations are, of course, less
desirable if we want to determine the absolute age of a star or
stellar population but they are very powerful if we want to know in
which order the stars formed and what time-scales were involved ,
i.e. the study of galaxy formation and evolution. The good thing with
the isochrone method is that it is, reasonably, straightforward to
derive the ages also for large samples of stars (but see
\cite[J{\o}rgensen \& Lindegren (2005)]{jl2005}) as well as for old
stars. The less useful aspect is that we are mainly limited to using
the turn-off stars.  For an older population this implies the
inherently faint, but numerous, F and  G type dwarf and sub-giant
stars. In order to construct the HR-diagram we need to know the
distances to the stars. This is difficult to do for large numbers of
stars once we are outside the volume covered by Hipparcos. However, it
is possible to derive the distance if the star is assumed to be a
dwarf or if the star can be determined to be a dwarf star. Str\"omgren
photometry and some other photometric systems are able to determine
the evolutionary state of a star. Some examples of how the Str\"omgren
photometry can be used to this end are given in \cite[Schuster et
al. (2006)]{sn2006}, \cite[von Hippel \& Bothun (1993)]{vonhippel},
and \cite[J{\o}nch-S{\o}rensen (1995)]{js1995}. So far these studies
have mainly been limited to the solar neighbourhood due to the
observational equipment available. Recent studies are trying to remedy
this situation by using CCD images obtained with wide-field
cameras. An early example is given in \cite[\'Arnadott\'ir et
al. (2008)]{aa2008}.

\section{The ages of the stellar disks}

The main tracers for age-dating the thin disk are open clusters, the
luminosity function of white dwarfs, and, recently,
nucleocosmochronology. Generally, estimates of the age of the thin
disk using the luminosity function of white dwarfs find a lower limit
for the age of around 9 Gyr (e.g.  \cite[Leggett et
al. (1998)]{legget98}, \cite[Knox et al. (1999)]{knox99}, and
\cite[Oswalt et al. (1995)]{oswalt95}). 

Open clusters indicate a similar lower age for the thin disk. It is
interesting to note the existence of open clusters that are both old
as well as metal-rich. NGC\,6791 has a metallicity of +0.35 dex and an
age between 8 and 9 Gyr, \cite[Grundahl et al. (2008)]{grundahl}. Such
old stars are normally not considered to be able to be that
metal-rich. In our new, local sample of stars we seem to pick up a few
metal-rich and old stars that have thin disk kinematics and thin disk
abundance patterns.

The thin disk hosts the majority of the younger stars.
In general young stars rotate more rapidly than older stars and they have
more chromospheric activity. As they grow older they rotate more
slowly and their outer atmospheres become less active.  These
characteristics can be utilised to estimate the age of a star,
\cite[Mamajek \& Hillenbrand (2008)]{mamajek2008} and \cite[Barnes
(2007)]{barnes2007}.  However, none of these measures are particularly
straightforward.  A recent example of how they could be combined in
order to give better age estimates is given by \cite[Mamajek \&
Hillenbrand (2008)]{mamajek2008}. They show that, with their new measure of
stellar ages, combined from rotation and activity measures, the star
formation history of the thin disk have been less variable than
previously thought.


The thick disk is in general found to be exclusively old.  The age
estimates for the thick disk have been done either by studying local,
kinematically selected samples or by studying the turn-off colour for
stars well above the galactic plane where the thick disk
dominates (typically about 1 kpc and higher, compare e.g. \cite[Gilmore
et al. (1995)]{gilmore95}). 

Recent studies of kinematically selected thick disk samples in the
solar neighbourhood appear to agree that the thick disk is old and,
essentially, all older than kinematically defined thin disk samples
(see e.g. \cite[Bensby et al. (2005)]{bensby2005} and \cite[Reddy et
al. (2006)]{reddy}). It is clear that the kinematic definitions only
are statistical and that we will never be able to create a sample that
is completely free from thin disk stars. It is especially important to
keep in mind that the young stars in the stellar disk (thin or
thick) have a rather lumpy distribution in velocity space.  This
enables the identification of stars that potentially have a common
origin but it complicates the division of stars into thin and thick
disk (for a recent discussion see \cite[Holmberg et
al. (2007)]{holmberg}). As shown in \cite[Holmberg et
al. (2007)]{holmberg}, as we progress to older stars the kinematics
change and the velocity distributions get smoother. This should not be
surprising as any older sample will be more dominated by the thick
disk, for which, not much lumpiness has been observed so far (but see
\cite[Gilmore et al. (2002)]{gilmore2002}, \cite[Schuster et
al. (2006)]{sn2006}, and \cite[Wyse et al. (2006)]{wyse2006} for
discussions of the last merger and how that has influenced the local
as well as not so local stellar kinematics). Not only the velocity
dispersions are important to consider but also how large a portion the
thick disk contributes in the solar neighbourhood (the normalization
of the stellar number density). In \cite[Bensby et
al. (2005)]{bensby2005} we show that our selection criteria are rather
robust against changes in this normalization. It remains to be fully
investigated how sensitive the selection criteria are to the presence
of lumpy velocity distributions.

In this context the study of volume-limited samples become
increasingly important.  \cite[Fuhrmann (2008)]{f08} studied a
volume-limited sample of stars within 25 pc from the sun. He
identifies the stars that are enhanced in [Mg/Fe] with the thick
disk. All of these stars are found to be older than the stars he
associate with the thin disk, but no specific ages are given. In the
next section we will revisit the volume-limited samples in comparison
to the samples selected based on kinematics.

\section{A new local sample of late F and early G dwarf stars -- the
local  disk(s) revisited}
\label{sect:new}

\begin{figure}[b]
\begin{center}
 \includegraphics[angle=-90,width=15cm]{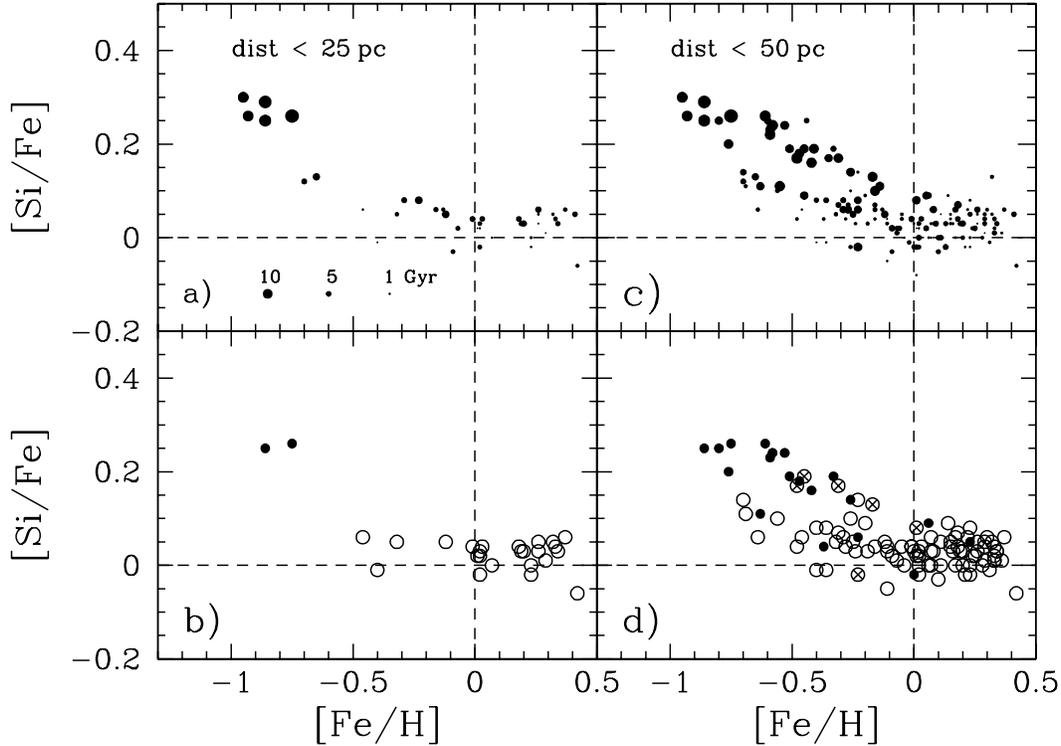} 
 \caption{[Si/Fe] vs. [Fe/H] for four sub-samples drawn from the full
   sample of Bensby et al. (2009 in prep). All stars shown have age
   determinations with relative errors less than 2 Gyr. Note that
   these are not volume-complete samples, only volume-limited.  {\bf
     a.} All stars from Bensby et al. (2009, in prep) within 25
   pc. The size of the symbols indicate their ages with larger symbols
   representing higher ages. The scale for the ages is indicated at
   the lower part of the panel. {\bf b.} A kinematically selected
   sub-set of the stars in a.  $\bullet$ marks stars that are ten
   times more likely to be thick than thin disk members and $\circ$
   marks stars that are ten times more likely to be thin than thick
   disk members.{\bf c.}  All stars from Bensby et al. (2009, in prep)
   within 50 pc. The size of the symbols indicate their ages with
   larger symbols representing higher ages. Same sizes are used as in panel
   a. {\bf d.} A kinematically selected sub-set of the stars in c.
   $\bullet$ marks stars that are ten times more likely to be thick
   than thin disk members and $\circ$ marks stars that are ten times
   more likely to be thin than thick disk members. Stars marked with
   an additional $\times$ are stars that are ten times more likely to
   be thin disk members than thick disk but also have an age larger
   than 8 Gyr. }
   \label{fig:sife}
\end{center}
\end{figure}

\begin{figure}[b]
\begin{center}
 \includegraphics[angle=-90,width=15cm]{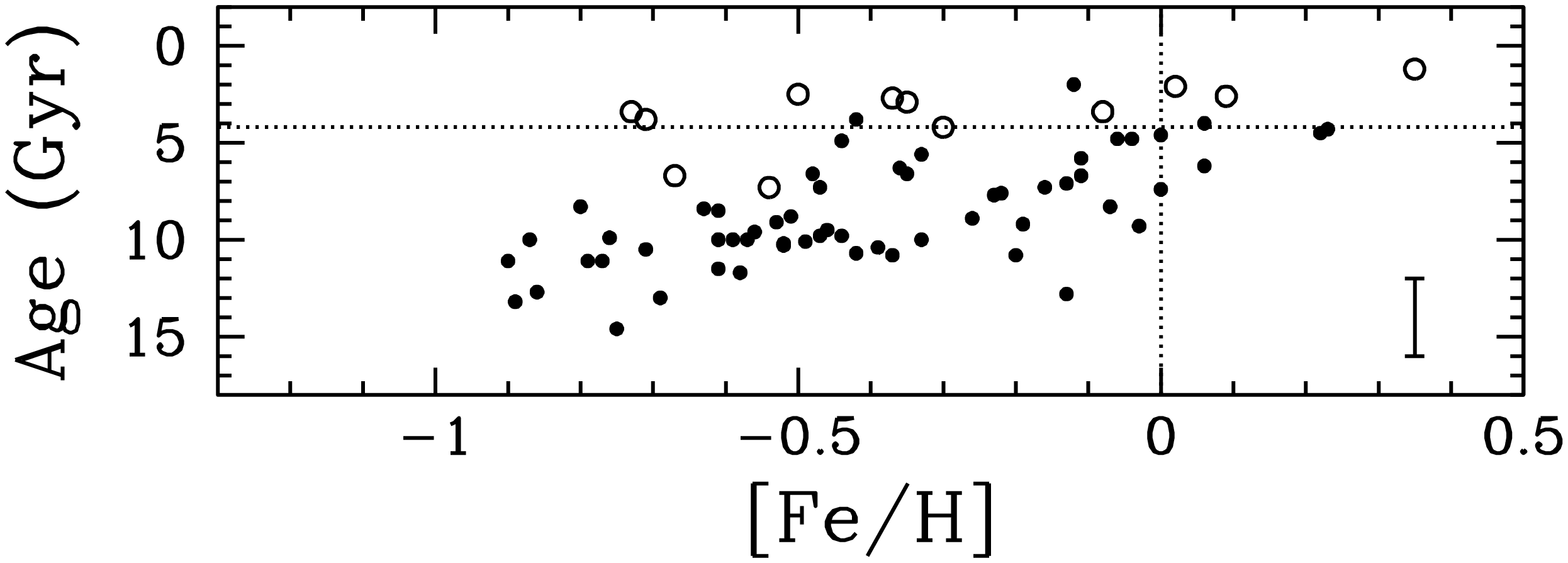} 
 \caption{Ages and metallicities for a sample of thick disk stars from
   our new study.  The stars shown all are ten times more likely to
   belong to the thick as opposed to the thin disk. The estimated
   error in the derived ages are less than 2 Gyr. The error-bar in the
   lower right hand corner shows a 2 Gyr error. Metallicities are
   based on spectroscopy. The position of the sun is marked by two
   dotted lines. $\alpha$-enhancement has been taken into account when
   determining the ages (see Sect.\ref{sect:new}). The filled circles
   show stars with effective temperatures less than 6000\,K and the
   open circles the 9 thick disk stars that have effective
   temperatures larger than 6000\,K.}
   \label{fig1}
\end{center}
\end{figure}

We have obtained high-resolution, high S/N spectra for about 900 dwarf
stars. The data have been obtained with several spectrographs but in
general S/N$>$250 and R$\geq$65,000 (apart from the sub-set of stars
originally observed with FEROS \cite[Bensby et al. (2003)]{bensby2003}
which have R=48,000). In \cite[Feltzing \& Bensby (2008)]{uppsala} we
presented the kinematic properties and some elemental abundances for a
sub-sample of about 550 F and G dwarf stars.

The ages for these stars have been derived using Yonseii-Yale
isochrones (see \cite[Bensby et al. (2005)]{bensby2005}) where we also
allow for enhancement in $\alpha$-elements. Taking the
$\alpha$-enhancement into account is important as, for a given star,
the age will be lower should it be enhanced in these elements as
opposed to if it is not. The effect of taking the $\alpha$-enhancement
into account is thus that any age-gap between the thick and the thin
disk decreases (the thin disk stars are not at all or only moderately
enhanced in $\alpha$-elements and thus there is only a small or no
effect on their ages when $\alpha$-enhancement is included in the
 age estimates).

 In Fig.\ref{fig:sife} a) and c) we are attempting a comparison with
 results from \cite[Fuhrmann (2008)]{f08} and show the stars within 25
 and 50 pc, respectively.  In the plots we have also coded the age of
 the stars such that an older star has a bigger symbol. It is clear
 from Fig.\,\ref{fig:sife} a) and c) that stars that are enhanced in
 an $\alpha$-element, here Si, are also older than stars that are not
 enhanced. On average, our volume-limited samples appear to show the
 same sort of trends that \cite[Fuhrmann (2008)]{f08} found. Although
 the sample within 25 pc is very small and incomplete, still stars
 enhanced in Si are old and the young stars are not enhanced and also
 show a tight trend of [Si/Fe] vs. [Fe/H]. Fig.\,\ref{fig:sife} c)
 further shows that there appear to be a real separation between the
 two trends, i.e. one for younger and one for older stars. It should
 be kept in mind that our sample is not volume-complete and also that
 \cite[Fuhrmann (2008)]{f08} imposes some further criteria on the
 stellar parameters for those stars that he includes in his final
 plots. For now, we are showing all stars, covering the full parameter
 space sampled within our programme (compare e.g.  \cite[Feltzing \&
 Bensby (2008)]{uppsala} and Bensby et al. in prep.).

Figures\,\ref{fig:sife} b) and d) then show the volume-limited
samples but with a kinematic selection imposed as well such that we
select stars that are ten times more likely to be thick disk than thin
disk to represent the thick disk and vice versa for the thin disk. It
is intriguing to see that the kinematically selected thin disk stars
mimics the trend found for the younger stars and the thick disk
mimics the trend found for the older stars. 

In \cite[Feltzing \& Bensby (2008)]{uppsala} we identified a small
number of stars on typical thin disk orbits but with enhanced
abundances for the $\alpha$-elements. These stars were found to be old
(older than about 8 Gyr in our determination). In Fig.\,\ref{fig:sife}
d) these stars are explicitly marked. For further discussion about
plausible origins for these stars we refer to \cite[Feltzing \& Bensby
(2008)]{uppsala}. It is worth nothing, however, that it is essentially
these stars that make the downward trend of [Si/Fe] in the
kinematically selected thick disk sample blend in with the thin disk
sample.

Figure\,\ref{fig1} shows the age-metallicity plot for a first
selection of stars with kinematics that make them very likely thick
disk candidates. All of these stars are ten times more likely to
belong to the thick than to the thin disk. As can be seen the bulk of
these stars are older than the sun and they have a mean age of around
10 Gyr. They cover that whole metallicity range from -1\,dex to
solar. For this first attempt at establishing if there is an
age-metallicity relation present in our kinematically defined sample
we have only included stars for which we could determine the ages to
better than 2 Gyr.  As our stars originally are essentially selected
only based on their kinematic properties and a metallicity estimated
from photometry we cover a reasonably large range of effective
temperatures. In Fig.\,\ref{fig1} we have chosen to show the stars
with effective temperatures larger than 6000\,K with a separate
symbol. Not surprisingly, these stars are in general young. If they
really belong to the thick disk then that would be rather challenging
for any of the models put forward for the formation of the thick
disk. However, our method to determine the stellar ages is ``simple''
and as these apparently young stars are in regions of the HR-diagram where the stellar
tracks show various ``kinks'' such a simple age estimate might go wrong in the
estimate of the error. We will therefore redo all
our ages using the method developed by \cite[J{\o}rgensen \& Lindegren
(2005)]{jl2005}. This method provides a better and more realistic
estimate of the error in the age determination. For now we would,
however, like to caution against over interpreting apparent young ages
present in kinematically defined thick disk samples.
 
\section{Summary}

Age determinations of the stellar disk(s) are inherently complicated.
There are several factors that makes it hard to define the age of
either disk, not the least the mixture of stellar populations in the
solar neighbourhood. The absolute ages of individual stars may be
obtained through e.g. astro-seismology and
nucleocosmochronology. However, for the study of the stellar
populations as such, fitting of isochrones to well defined samples and
the fitting of the white dwarf luminosity function using cooling
tracks might be more appropriate.

In most current studies stars with kinematics typical of the thick
disk are, on average, found to be older than stars with kinematics
typical of the thin disk. There appear to be an age-metallicity
relation present in the thick disk.  This is found in studies using
various technqieus. However, the exact definition of the thick disk in
relation to the thin disk in terms of stellar kinematics is not
straightforward and will need more work.

\begin{discussion}

\end{discussion}

\end{document}